\documentclass[twocolumn,showpacs,preprintnumbers,amsmath,amssymb,prl]{revtex4}
\usepackage{graphicx}
\usepackage{amssymb}
\usepackage{amsmath}
\usepackage{dcolumn}
\usepackage{bm}
\include{natbib}

\begin{document}
\title{Space-time resolved electrokinetics in cylindrical and semi-cylindrical microchannels}
\author{Michele Campisi}
\affiliation{Department of Physics,University of North Texas, P.O.
Box 311427, Denton, TX 76203-1427,USA}
\date{\today}
\begin{abstract}
It is shown show how to employ Bessel-Fourier series in order to
obtain a complete space-time resolved description of
electrokinetic phenomena in cylindrical and semi-cylindrical
microfluidic channels.
\end{abstract}

\pacs{
47.15.-x, 
47.65.+a, 
73.30.+y, 
82.39.Wj. 
}

\keywords{Lab-on-a-chip, Microfluidics, Micro Total Analysis
System, Electrokinetics, Non-equilibrium thermodynamics, Onsager
relations}

\maketitle

The employment of Fourier series and Laplace transforms allows to
solve exactly the low Reynolds number Navier-Stokes equation in
rectangular microchannels \cite{Campisi05,Campisi06}. In Refs.
\cite{Campisi05,Campisi06} this has been used to find the
space-time resolved velocity field and current density generated
by the sudden application of a possibly time dependent pressure
gradient and/or an electric field. This allowed to fully
characterize transient and steady state electrokinetic phenomena
in rectangular microchannels. A similar characterization in the
case of cylindrical and semi-cylindrical is still missing. These
channel geometries (particularly the latter) are relevant for
microfluidics devices fabricated from glass substrates. Typically
microchannels are obtained from such substrates by means of
isotropic etching which produces semi-cylindrical grooves. Here we
shall show how to adapt the mathematical methods presented in Ref.
\cite{Campisi05,Campisi06} to the cylindrical and semi-cylindrical
geometries, in order to allow for a space-time resolved
description of electrokinetic phenomena in these cases. The
physics involved will be qualitatively the same as in the
rectangular geometry, thus we will not delve into a detailed
discussion, but present only the mathematical method.

\textbf{Cylindrical geometry} As reported before \cite{Campisi05}
the starting point for studying space-time resolved electrokinetic
phenomena in microchannels is the low Reynolds number
incompressible Navier-Stokes equation with a generic time and
space dependent body force. The natural set of coordinates for
solving this problem within cylindrical microchannels is the
cylindrical coordinates $(r,\varphi,x)$, where $x$ runs along the
channel axis and $r,\varphi$ are polar coordinates in the channel
cross-section. Using this coordinate set, the $x$ component of the
Navier-Stokes equation takes the following form
\begin{equation}\label{eq:NS-general}
    \frac{\partial u(r,\varphi,t)}{\partial t}-\nu \left({\partial^2 \over \partial r^2} + {1 \over r}
{\partial \over \partial r} + {1 \over r^2} {\partial^2 \over
\partial \varphi^2} \right) u(r,\varphi,t)= f(r,\varphi,t)
\end{equation}
where $u$ is the $x$ component of the velocity field and $f$ is
the body force. In the case of rectangular geometry the solution
of the Navier-Stokes equation was found by means of double Fourier
series expansion. In other words the set of functions $\psi_{k,q}=
sin\left(\frac{k\pi y}{2H}\right)sin\left(\frac{q\pi
z}{2W}\right)$, $k,q\in \mathbb{N}_+$ was chosen as the
appropriate complete basis set for the solution of the problem
with null boundary condition over a rectangular boundary.
Likewise, for a circular boundary of radius $R$ the appropriate
basis set is given by the following functions
\begin{eqnarray}\label{eq:basis-set}
    \psi_{m,k}(r,\varphi)=e^{im\varphi}J_m(\alpha_{m,k}r/R) & & m \in \mathbb{Z}, k \in \mathbb{N}_+ \nonumber\\
\end{eqnarray}
The symbol $J_m$ denotes the $m^{th}$ Bessel function of the first
kind, and $\alpha_{m,k}$ denotes its $k^{th}$ zero. For sake of
completeness let us recall that the $J_m$ are defined as the
solutions of the equations
\begin{equation}\label{eq:bessel-equation}
    \rho^2{\partial^2 \over \partial \rho^2}J_m(\rho) + \rho {\partial \over
    \partial \rho}J_m(\rho) + (\rho^2-m^2)J_m(\rho)=0
\end{equation}
and that for fixed $m$ the following orthogonality relation exists
between the functions $J_m$:
\begin{equation}\label{eq:bessel-ortho}
    \int_0^1 d\rho \rho
    J_m(\alpha_{m,k}\rho)J_m(\alpha_{m,q}\rho)= {1 \over 2}
    \delta_{k,q} [J_m(\alpha_{m,k})]^2
\end{equation}
In a similar fashion the $y_m(\varphi)=e^{im\varphi}$ are
solutions of $y'=imy$ and obey the orthogonality condition:
\begin{equation}\label{eq:exp-ortho}
    \int_0^{2\pi} d\varphi y_m^*(\varphi) y_n(\varphi) = 2\pi \delta_{m,n}
\end{equation}
where the symbol ``$*$'' denotes complex conjugation. Using Eq.
(\ref{eq:bessel-ortho}) and Eq. (\ref{eq:exp-ortho}), allows to
normalize the basis set (\ref{eq:basis-set}) and obtain the the
following complete orthonormal basis:
\begin{equation}\label{eq:basis-set-normalized}
    \psi_{m,k}(r,\varphi)={e^{im\varphi}J_m(\alpha_{m,k}r/R) \over \sqrt{\pi}R J_{m+1}(\alpha_{m,k})}\quad m \in \mathbb{Z}, k \in \mathbb{N}_+
\end{equation}
\begin{equation}\label{}
    \int_0^{2\pi} d\varphi \int_0^{R} rdr
    \psi_{m,k}^*(r,\varphi)\psi_{n,q}(r,\varphi)=\delta_{m,n}\delta_{k,q}
\end{equation}
The completeness of the set (\ref{eq:basis-set-normalized}) allows
to expand the functions $f$ and $u$ in a double generalized
Fourier series (the Bessel-Fourier series), as follows:
\begin{equation}\label{}
    u(r,\varphi,t)= \sum_{m \in \mathbb{Z}} \sum_{k \in
    \mathbb{N}_+} u_{m,k}(t) \psi_{m,k}(r,\varphi)
\end{equation}
with coefficients given by:
\begin{equation}\label{}
    u_{m,k}(t)=\int_0^{2\pi} d\varphi \int_0^{R} rdr
    \psi_{m,k}^*(r,\varphi)u(r,\varphi,t)
\end{equation}
Expanding the Navier-Stokes equation (\ref{eq:NS-general}) over
the basis (\ref{eq:basis-set-normalized}) gives, thanks to the
property (\ref{eq:bessel-equation}) and in complete analogy with
what was found previously for the rectangular geometry
\cite{Campisi05}, the following set of equations:
\begin{equation}\label{eq:NS-mk}
    \frac{\partial}{\partial t}
    u_{m,k}(t)+\nu \Delta_{m,k}^2u_{m,k}(t)=f_{m,k}(t)
\end{equation}
with null-boundary condition over the $r=R$ circumference
automatically fulfilled. The quantities $\Delta_{m,k}^2$, given by
the following formula,
\begin{equation}\label{}
    \Delta_{m,k}^2={\alpha_{m,k}^2 \over R^2}
\end{equation}
are, so to speak, the expansion coefficients of the Laplacian
operator over the given basis set (\ref{eq:basis-set-normalized}).
The $f_{m,k}(t)$ represent the expansion coefficients of the body
force $f$:
\begin{equation}\label{}
    f_{m,k}(t)=\int_0^{2\pi} d\varphi \int_0^{R} rdr
    \psi_{m,k}^*(r,\varphi)f(r,\varphi,t)
\end{equation}
If the liquid is considered as initially at rest Eq.
(\ref{eq:NS-mk}) must be solved with for the initial conditions
$u_{m,k}(t=0)=0$. Following the general line drawn in Ref.
\cite{Campisi05}, the solution is easily expressed in the Laplace
space as:
\begin{equation}\label{eq:relation u(s) f(s)}
    \widetilde{u}_{m,k}(s)=\frac{\widetilde{f}_{m,k}(s)}{s+\nu
    \Delta_{m,k}^2}.
\end{equation}
which expresses the Laplace transform of each coefficient of the
velocity profile in terms of the corresponding Laplace transformed
components of the driving force. The solution $u(r,\varphi,t)$ is
obtained by anti-Laplace-transform ($\mathcal{L}^{-1}$) and
summing up:
\begin{equation}\label{eq:formal-solution}
    u(r,\varphi,t)=\sum_{m \in \mathbb{Z}} \sum_{k \in
    \mathbb{N}_+}\mathcal{L}^{-1}\left[\frac{\widetilde{f}_{m,k}(s)}{s+\nu
    \Delta_{m,k}^2}\right]\psi_{m,k}(r,\varphi)
\end{equation}
For a pressure driven flow the body force $f$ would be given by
$f(r,\varphi,t)=\frac{\Delta P(t)}{\rho L}$, with $L$ the length
of the channel, $\rho$ the liquid density and $\Delta P(t)$ the
possibly time dependent pressure difference applied at the ends of
the channel. For an electro-osmotically driven flow the body force
would be given by $\frac{\rho_e(r,\varphi)}{\rho}E(t)$, with
$E(t)$ the applied electric field and $\rho_e$ the electric double
layer (EDL) charge density that spontaneously forms at the solid
liquid interface. The latter can be found by solving the
Poisson-Boltzmann equation \cite{Campisi05} for the electric
double layer potential $\psi$ within the Debye-H\"{u}ckel
approximation \cite{Hunter}:
\begin{equation}\label{eq:linearPoissonBoltz}
    \left({\partial^2 \over \partial r^2} + {1 \over r}
{\partial \over \partial r} + {1 \over r^2} {\partial^2 \over
\partial \varphi^2} \right) \psi (r,\varphi)= \chi ^2 \psi(r,\varphi),
\end{equation}
where $\chi$ is the inverse Debye length. Expanding over the basis
(\ref{eq:basis-set-normalized}), and using the Poisson equation
$\Delta\psi=-\rho_e/\varepsilon$, like in \cite{Campisi05} one
obtains the cherge density coefficients:
\begin{equation}\label{eq:kq-charge-density}
    \rho_{e(m,k)}=-\varepsilon \zeta \chi^2 \frac{\Delta_{m,k}^2 I_{m,k}}{\Delta_{m,k}^2+\chi ^2}
\end{equation}
where $I_{m,k}$ denote the expansion coefficients of the unity:
\begin{equation}\label{}
    I_{m,k}(t)=\int_0^{2\pi} d\varphi \int_0^{R} rdr
    \psi_{m,k}^*(r,\varphi).
\end{equation}
The solution of the problem in the cylindrical geometry is
formally equivalent to that of the rectangular geometry. The only
difference is contained in the way the expansion coefficeint are
calculated: using the double Fourier series for the rectangular
case, and using the Bessel-Fourier series in the cylindrical case.
Thus once the basis set appropriate for a given geometry is found,
the problem is automatically solved.

\begin{figure}
  \includegraphics[width=8cm]{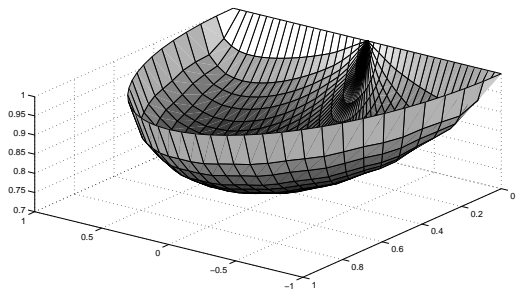}
  \caption{Typical EDL charge density profile in non-dimensional units ($R=1, \rho_e(1,\phi)=1$).
  The first $100 \times 100$ Bessel-Fourier coefficients have been employed to generate the plot.}
  \label{fig:charge-density}
\end{figure}

\textbf{Semi-cylindrical geometry} In this case, the function
$u(r,\varphi)$ must obey not only the condition of being null for
$r=R$, but also for $\varphi=0,\pi$. Seen in a different way, the
function $u$  must be odd under the operation $\varphi \rightarrow
- \varphi$. Therefore its expansion series would contain only odd
terms, i.e., it would be of the type:
\begin{equation}\label{}
    \sum_{m \in \mathbb{Z}} \sum_{k \in
    \mathbb{N}_+} \left[\psi_{m,k}(r, \varphi)-\psi_{-m,k}(r,\varphi)\right]
\end{equation}
Namely it would contain only sine terms. Therefore the orthonormal
basis set suitable for the semi-cylindrical geometry is:
\begin{equation}\label{eq:basis-semi-cyl}
    \phi_{m,k}(r,\varphi)={2\sin(m\varphi)J_m(\alpha_{m,k}r/R)
    \over \sqrt{\pi}R J_{m+1}(\alpha_{m,k})}\quad m, k \in \mathbb{N}_+
\end{equation}
Where the $\phi_{m,k}$ satisfy the orthonormality conditions:
\begin{equation}\label{}
    \int_0^{\pi} d\varphi \int_0^{R} rdr
    \phi_{m,k}(r,\varphi)\phi_{n,q}(r,\varphi)=\delta_{m,n}\delta_{k,q}
\end{equation}
We will write the expansion of $u$ as:
\begin{equation}\label{}
    u(r,\varphi,t)= \sum_{m ,k \in
    \mathbb{N}_+} u'_{m,k}(t) \phi_{m,k}(r,\varphi)
\end{equation}
with coefficients given by:
\begin{equation}\label{}
    u_{m,k}'(t)=\int_0^{\pi} d\varphi \int_0^{R} rdr
    \phi_{m,k}(r,\varphi)u(r,\varphi,t)
\end{equation}
where the prime symbol is used to distinguish these coefficients
from those defined previously. Adopting the same notation for the
expansion of the body force, again the expansion of the
Navier-Stokes equation leads to the solution
\begin{equation}\label{}
    \widetilde{u}_{m,k}'(s)=\frac{\widetilde{f}_{m,k}'(s)}{s+\nu
    \Delta_{m,k}^2}.
\end{equation}
which is formally equivalent to Eq. (\ref{eq:relation u(s) f(s)}).
The charge density will be given by
\begin{equation}\label{eq:kq-charge-density'}
    \rho'_{e(m,k)}=-\varepsilon \zeta \chi^2 \frac{\Delta_{m,k}^2 I'_{m,k}}{\Delta_{m,k}^2+\chi ^2}
\end{equation}
where
\begin{equation}\label{}
    I'_{m,k}(t)=\int_0^{\pi} d\varphi \int_0^{R} rdr
    \phi_{m,k}(r,\varphi)
\end{equation}
which is formally equivalent to Eq. (\ref{eq:kq-charge-density}).
As an illustration of the method Fig. \ref{fig:charge-density}
shows a typical plot of the EDL charge density obtained from Eq.
(\ref{eq:kq-charge-density'}). All the information relevant for
the description of electrokinetic phenomena in cylindrical and
semi-cylindrical microchanels is contained in the coefficients
$u_{m,k},\rho_{e(m,k)}$ and $u'_{m,k},\rho'_{e(m,k)}$,
respectively. These can be used like in Ref.
\cite{Campisi05,Campisi06} to obtain a space-time resolved
description of electrokinetic phenomena. For example, for the
semi-cylindrical geometry one finds the following generalized
conductance matrix:
\begin{equation}\label{eq:conductance-matrix-explict}
    \mathbf{M} =
    \frac{1}{\rho L}
    \sum_{m,k}\frac{e^{- i\theta'_{m,k}(\omega)}}{\sqrt{\omega^2+\nu^2 \Delta_{m,k}^4}}
    \mathbf{A'}_{m,k}
\end{equation}
Where $L$ is the channels's length, $\omega$ is the angular
frequency of the driving body force,
\begin{equation}
\mathbf{A'}_{m,k} \doteq
\left(\begin{array}{cc} I'^2_{m,k}            & I'_{m,k}\rho'_{e(k,q)}  \\
                        I'_{m,k}\rho'_{e(k,q)} & \rho'^2_{e(k,q)}
      \end{array}
\right)
\end{equation}
and
\begin{equation}\label{eq:def-theta}
    \theta'_{m,k}(\omega)=\arctan\left(\frac{\omega}{\nu
\Delta^2_{m,k}}\right).
\end{equation}

\textbf{Acknowledgements} The coefficients $\alpha_{m,k}$ used to
produce the plot in Fig. \ref{fig:charge-density} have been
calculated numerically with the open-access \texttt{MATLAB}
program \texttt{besselzero.m} written by Greg von Winckel.


\newpage
\end{document}